\documentstyle[12pt]{article}

\def\mxth{\mathsurround=0pt }
\def\xversim#1#2{\lower2.pt\vbox{\baselineskip0pt \lineskip-.5pt
  \ialign{$\mxth#1\hfil##\hfil$\crcr#2\crcr\sim\crcr}}}
\def\simgr{\mathrel{\mathpalette\xversim >}}
\def\simle{\mathrel{\mathpalette\xversim <}}
\newcommand{\for}{\mbox{  for $i$ = 1, 2, 3,}}
\newcommand{\GeV}{\mbox{  GeV}}
\newcommand{\doto}{\mbox{o}}

\begin{document}
\title{COUPLING CONSTANT
AND YUKAWA COUPLING  UNIFICATIONS:
UNCERTAINTIES AND CONSTRAINTS}

\author{Nir Polonsky\footnote{}\\
{\it Department of Physics, University of Pennsylvania}\\
{\it Philadelphia, Pennsylvania, 19104, USA}\\}
\date{UPR-0571T}

\maketitle
\setlength{\baselineskip}{2.6ex}
\begin{abstract}
The status of coupling constant unification (with and without
a unification of Yukawa couplings) is discussed.
Uncertainties associated with the input coupling constants,
$m_{b}$ and $m_{t}$, threshold corrections at the low and high scale,
and possible nonrenormalizable operators are described
and a discrepancy between effective and physical scales
 is pointed out.
Theoretical uncertainties in the predictions of
$\alpha_{s}(M_{Z})$, $m_{b}$,
and the unification scale, $M_{G}$, are discussed and estimated.
Constraints on the
super-partner spectrum are found to be weak if
uncertainties
associated with the high-scale are included.
However, requiring $h_{b} = h_{\tau}$
at $M_{G}$ excludes $3 \simle \tan \beta \simle 40$
(for $m_{t} \simle 200$ GeV).
\end{abstract}

The standard model (SM) couplings were recently shown\cite{firsts}
to meet at a point, $M_{G} \approx 10^{16} -
10^{17}$  GeV, when extrapolated to high energy assuming
a grand desert and the spectrum of the minimal supersymmetric
standard model
(MSSM) above the weak scale (a minimal one-step scenario).
This is still true if one admits
additional $U_{1}$ factors or a small number of complete
multiplets of a gauge group of some grand unified theory (GUT).
(The latter
affect the predictions only at the two-loop level.) Otherwise,
it would be difficult to
relax the assumption on either the spectrum or the desert scenario
without relaxing the other, i.e., introducing
additional matter at some intermediate scales. In order to maintain
the predictability of the model, we assume
hereafter a  minimal one-step scenario.

 For certain values of the $t$-quark (pole) mass, $m_{t}$, and of
the two Higgs doublet expectation value ratio, $\tan \beta$, Yukawa
couplings of the third family fermions also unify at $\sim M_{G}$
(when extrapolated under the same assumptions)\cite{yukfirsts},
i.e., $h_{b} = h_{\tau}$ ($ = h_{t}$),
as is implied by certain $SU_{5}$ ($SO_{10}$) and similar GUT's.
(One usually assumes that some perturbation
modifies the couplings or the masses
of the two light families where, in principle, similar
relations should, but do not, hold.)
While the coupling constant extrapolation is decoupled
to a good approximation from
that of the Yukawa sector, the latter is controlled
by the coupling constants. It is the balance between the
coupling constants
and the Yukawa couplings
that determines the infra-red fixed points
in the Yukawa coupling renormalization flow.
We will examine the status of  coupling constant
unification  first\cite{us1}.
Then, we will further assume Yukawa unification at $\sim M_{G}$, and use
that assumption to constrain the $m_{t} - \tan \beta$ plane\cite{us2}.
While $m_{t}$
effects can
be treated as a correction term in the former case,
$m_{t}$ is a free parameter in the latter. $\alpha_{s}$
will be fixed by the unification, and as we point out below,
is thus a quadratic function
of $m_{t}$.

The naive scenario has to be somewhat relaxed in order to obtain a more
realistic picture. Below we will perturb the unification and desert
assumptions by taking $m_{t} > M_{Z}$, an arbitrary sparticle spectrum
(below the TeV scale), and by
considering a split spectrum and nonrenormalizable
operators (NRO's) at the high-scale. For the latter, we will assume
a minimal $SU_{5}$ GUT. We do so in order to
explicitly realize the magnitude
of the effects and the role of the
constraints coming from proton decay at the
loop-level\cite{proton};
however, this can be easily generalized\cite{us1}.
(Having larger GUT gauge groups does not imply larger corrections.)
We assume that all the corrections are
consistent with perturbative treatment,
and that there is no conspiracy among the different correction terms.
We will add theoretical uncertainties in quadrature as a guideline only.

Let us then write
\begin{equation}
\frac{1}{\alpha_{i}(M_{Z})} = \frac{1}{\alpha_{G}}
+ b_{i}t + \theta_{i} + H_{i} - \Delta_{i} \,\, \for
\label{eq1}
\end{equation}
where we can neglect the two-loop
contribution from the Yukawa sector, $H_{i}$.
All other uncertainties and corrections (including conversion to
$\overline{DR}$) are included in the correction
functions $\Delta_{i}$. $\alpha_{G}$ is
the coupling at $M_{G}$
and $t \equiv \frac{1}{2\pi}\ln \frac{M_{G}}{M_{Z}}$
is a convenient parametrization of the unification point, $M_{G}$.
$b_{i}$ are the respective one-loop
$\beta$-function coefficients and $\theta_{i}$
are the two-loop corrections. One can then get expressions for
$\alpha_{G}$, $t$, and either $\alpha_{s}(M_{Z})$ or the weak angle,
$s^{2}(M_{Z})$, in terms of only $\alpha(M_{Z})$ and either
$s^{2}(M_{Z})$ or $\alpha_{s}(M_{Z})$. $\theta_{i}$ are calculated
iteratively and $\Delta_{i}$ determine the theoretical uncertainties.

$s^{2}(M_{Z})$  is strongly correlated with $m_{t}$. It is then useful
to define a ($\overline{MS}$) $m_{t}$-independent quantity\cite{fit},
$s^{2}_{0} = 0.2324 \pm 0.0003$;
i.e., $s^{2}(M_{Z})$  obtained for a fixed $m_{t_{0}}$ ($= 138$ GeV),
and use $s^{2}_{0}$ to predict $\alpha_{s}(M_{Z})$. One can then account
for $m_{t} \neq 138$ GeV (i.e., $113 < m_{t} < 159$ GeV
from precision electroweak data) by including the leading (quadratic)
$m_{t}$ dependence in the correction functions $\Delta_{i}$\cite{us1}.
(This is accurate up to small logarithmic corrections). By following the
above procedure, (leading) $m_{t}$ effects are treated consistently
and we bypass the $\sim 8\%$ input uncertainty in the
$\alpha_{s}(M_{Z}) = 0.12 \pm 0.01$ range extracted from experiment.
The $\pm 0.01$ input uncertainty would
have induced  an uncertainty comparable
to the theoretical ones if we were using $\alpha_s$ to predict $s^{2}$.
Setting $\Delta_{i} \equiv 0$, as in the naive calculation, we obtain
$\alpha_{G}^{-1} = 23.41 \pm 0.04$, $t = 5.30 \pm 0.01$, and
$\alpha_{s}(M_{Z}) = 0.125 \pm 0.001$.
Turning on $m_{t} > M_{Z}$ correlates
$\alpha_{s}(M_{Z})$ with $m_{t}$, an effect that has to be taken into
account when, e.g., constraining the $m_{t} - \tan \beta$ plane.

Aside from $m_{t}$ effects, $\Delta_{i}$ consists
also of one-loop threshold
corrections (which is consistent with a two-loop calculation) and NRO
effects. The former have to be accounted
for at both scales\cite{bh},
and analytic treatment is then more instructive and convenient.
NRO that renormalize (and thus split) the couplings at $M_{G}$
are suppressed by the ratio of
$M_{G} \simgr 10^{16}$ GeV over some larger scale.
Even so, the effect can be  comparable to two-loop
effects\cite{us1,hs}. We parametrize
the $\Delta_{i}$ functions in terms of
7 effective parameters
(aside from $m_{t}$)
corresponding to the  sparticle and Higgs
doublet spectrum ($M_{1}$, $M_{2}$, $M_{3}$); the high-scale thresholds
($M_{V}$, $M_{24}$, $M_{5}$, for the heavy components
of the vector, adjoint and complex Higgs supermultiplets,
respectively); and an effective
NRO strength, $\eta$ (which can have either sign).
All these are described in detail in Ref. 3. The corrections
to the $\alpha_s(M_{Z})$ prediction due to each of these parameters
are illustrated in Figure 1.
The corrections are all comparable and have no fixed sign, i.e.,
\begin{equation}
\alpha_{s}(M_{Z}) \approx 0.125
\pm 0.001 \pm 0.005 ^{+ 0.005}_{-0.002} \pm{0.002} \pm 0.006,
\label{eq2}
\end{equation}
or $\alpha_{s}(M_{Z}) \approx 0.125 \pm 0.010$,
which is in good agreement with the data. The uncertainties quoted
in (\ref{eq2}) correspond to those of $\alpha$
and $s^{2}_{0}$, reasonable
choice of ranges of sparticle (and Higgs) spectra, high-scale spectra,
$113 < m_{t} < 159$ GeV, and to NRO effects, respectively.
They are intended to serve as an order of magnitude estimate only.
(Proton decay constrains $M_{5}$ and thus constrains the
high-scale threshold
negative contribution to the uncertainty, unless
one turns to simple extentions
or eliminates the colored triplet Higgs from the spectrum.)

To further illustrate the underlying formalism, let us discuss\cite{us1} in
greater detail the parameters $M_{i}$. They are defined as weighted sums,
\begin{equation}
\sum_{\zeta}\frac{b_{i}^{\zeta}}{2\pi}\ln\frac{M_{\zeta}}{M_{Z}}
\equiv \frac{b_{i}^{MSSM} -
b_{i}^{SM}}{2\pi}\ln\frac{M_{i}}{M_{Z}} \,\, \for
\label{eq3}
\end{equation}
where the summation is
over all the sparticles and the heavy Higgs doublet
and $b_{i}^{\zeta}$ is the $\zeta$-particle contribution to the respective
$\beta$-function.
$M_{i}$ determines the low-scale threshold  contribution
to $\Delta_{i}$ and it is a straight-forward
exercise to calculate it for any spectrum\cite{pok}.
(Similar model independent
parameters can be defined at the high-scale\cite{us1}.)
The parameters are sensitive to
the split between, for example, colored and uncolored sparticles.
$\Delta_{\alpha_{s}}$ is proportional to the combination
$25\ln\frac{M_{1}}{M_{Z}} -100\ln\frac{M_{2}}{M_{Z}}
+56\ln\frac{M_{3}}{M_{Z}}$
($\equiv  -19\ln\frac{A_{SUSY}}{M_{Z}}$),
and is potentially large and negative;
e.g., for a spectrum degenerate at $M_{SUSY}$.
However, in supergravity-inspired
MSSM $M_{3} \sim M_{gaugino}$; $M_{1} \sim m_{0},
\mu$; and $M_{2} \sim \mu$.
Thus, in general $M_{2} \simle M_{1}$ and $M_{2} \simle M_{3}$, and the
correction can be positive, depending on the split.
Rather than evaluating
$\Delta_{\alpha_{s}}$ in terms of a common mass, $M_{SUSY}$,
we can invert the logic and use
the above expression to define an effective scale, $A_{SUSY}$,
which is the relevant scale in the
problem. $A_{SUSY}$ can be as low as a few GeV
and does not contain by itself any
physical information (nor should we expect it to).
One should therefore calculate
$M_{1}$, $M_{2}$ and $M_{3}$, which determine also the
low-scale corrections to
$\alpha_{G}$ and $t$, and contribute to the corrections to
$m_{b}$. It is always possible to define
an effective scale in the context of a specific
prediction, but unlike the $M_{i}$ parameters, such scales
do not have an obvious  physical interpretation.

Similarly to (\ref{eq2}) we have
(using the same ranges for the parameters)
\begin{equation}
t \approx 5.30 \pm 0.01 \pm 0.09 ^{+ 0.31}_{-0.01} \pm0.02 \pm 0.025,
\label{eq4}
\end{equation}
which corresponds to $10^{16} \simle M_{G} \simle 2\times 10^{17}$ GeV.
$t$ is insensitive to
$\eta$ (i.e., to NRO's) but increases  significantly
if the heavy color octet and $SU_{2}$-triplet
Higgs coming from the adjoint
representation ($M_{24}$ in the above notation) are somewhat
lighter than the other heavy thresholds.
One can then increase the predicted $M_{G}$
up to $\sim 5 \times 10^{17}$ GeV while
maintaining a successful prediction of $\alpha_{s}$ by postulating
a large (but still consistent with a perturbative treatment)
and negative $\eta$. (The above considerations relax
proton-decay constraints\cite{proton}.)
However, it is not straight-forward to realize
such an interplay in  string-inspired GUT-like models
that, in general, do not admit adjoint and other large representations.

Finally, we require also $h_{b}(M_{G}) = h_{\tau}(M_{G})$. There is no
expression similar to (\ref{eq1})
for the (two-loop) Yukawa couplings and
one has to turn to numerical integration. By solving
a set of six two-loop renormalization group equations\cite{bj}
(neglecting the two light family Yukawa couplings and flavor mixings)
we can predict the $b$-quark (current) mass, $m_{b}$,
as a function of $m_{t}$ and $\tan \beta$. The leading dependences on
$m_{t}$ are from $h_{t}$ and the $\alpha_{s}(M_{Z})$ prediction
(the $s^{2}(M_{Z})$ input). It is useful to define
$m_{b} = \rho^{-1} \times m_{b}^{0}$, where $m_{b}^{0}$ is calculated to
two-loop (numerically) using the ideal desert and unification assumptions
(aside from $m_{t}$ effects). $\rho^{-1}$ is a multiplicative correction
function (e.g., the equivalent of the additive $\Delta_{\alpha_{s}}$,
only that $\Delta_{\alpha_{s}}$ in Eq. (\ref{eq2}), for example,
includes $m_{t}$ corrections and $\rho^{-1}$ does not).
The correction function, $\rho^{-1}$,  consists of corrections
to $\alpha_{s}(M_{Z})$, $\alpha_{G}$,
and $M_{G}$ ($t$) described previously,
as well as threshold effects that correct the running of the Yukawa
couplings either directly (by modifying
the respective $\beta$-function) or via
the modified running of the coupling constants
(and in particular, $\alpha_{3}$).
One could also allow a small
arbitrary split between the two unification points.
All these are described in detail in Ref. 4,
where we studied the effects using (approximate) analytic
correction expressions. We obtain $\rho^{-1} \approx 1 \pm 0.15$,
and thus, the constraint
$0.85m_{b}^{0}(5$ GeV) $< 4.45$ GeV.  The $0.85$
correction factor is estimated for a conservative choice
of ranges for the various parameters, and
is more sensitive to high-scale corrections to
the coupling constant unification than to the details
of the sparticle spectrum (for sparticles below the TeV scale).
We also use a conservative estimate
of the upper bound on the
current mass.

That constraint, in addition to  requiring a perturbative
Yukawa sector ($h < 3$ up to $M_{G}$), allows only
two branches in the $m_{t} - \tan\beta$ plane,
as illustrated in Figure 2.
The lower and upper bounds on $\tan\beta$ and the upper bound
on $m_{t}$
are from perturbative consistency
(i.e., are determined by the fixed-points).
The width of each branch is
determined by the $0.85$ correction factor
and the current mass upper bound.
$m_{t}$ correlations
increase the upper bound on $m_{t}$
($\sim 215 \pm 10$ GeV),
and as a result, the allowed area is  not sensitive
to $m_{t} \simle 200$ GeV. For $m_{t} \simgr 200$ GeV
some intermediate values of $\tan\beta$ are allowed.
Otherwise, $\tan\beta$ is strongly constrained,
i.e. $0.6 \simle \tan\beta \simle 3$ or $40 \simle \tan\beta \simle 60$.
The latter may be further excluded for $m_{t} \simle 170 \pm 10$ GeV
by requiring radiative breaking of $SU_{2} \times U_{1}$.

We can write an explicit constraint
on the sparticle (and Higgs doublet) spectrum
from coupling constant unification, i.e.,
\begin{equation}
0.110 \simle \alpha_{s}^{0}(M_{Z}) +
\frac{(\alpha_{s}^{0}(M_{Z}))^{2}}{28\pi}
\left[ 25\ln\frac{M_{1}}{M_{Z}} - 100\ln\frac{M_{2}}{M_{Z}}
+ 56\ln\frac{M_{3}}{M_{Z}}\right] \pm 0.008 \simle 0.130,
\label{c1}
\end{equation}
where $\alpha_{s}^{0}$ here is the
prediction for  ideal desert and unification
(i.e., $0.125$). Requiring in addition $h_{b}(M_{G}) = h_{\tau}(M_{G})$
implies a similar, but weaker, constraint which depends also on
$\max{(M_{gluino},M_{squark})}$, $\min{(M_{gluino},M_{squark})}$,
and on the heavy Higgs doublet mass, $M_{H}$.
(See Ref. 4.)
High-scale effects weaken the constraints significantly.

To summarize, coupling constant unification agrees very well
with the data. Yukawa coupling unification
strongly constrains $\tan\beta$ (independent of $m_{t} \simle 200$ GeV),
and is in agreement with the data
only in a small area of the $m_{t} -\tan\beta$ plane.
The above analysis may be further incorporated, e.g., in a one-loop
sparticle spectrum analysis\cite{us3},
and, in principle,
can give hints on the structure of the high-scale physics
(e.g., here we showed that $\tan\beta \approx 3 - 40$ will
disfavor a large class of models).

\subsection*{Acknowledgments}
This work was done with Paul Langacker, and was supported by the
Department of Energy Grant No. DE-AC02-76-ERO-3071.

 \newpage
 \begin{figure}
 \caption{Contributions of individual correction terms -- the SUSY
 effective mass parameters $M_{i}$; the heavy thresholds at the high-scale;
 the $t$-quark mass;
and NRO's at the high-scale -- to the $\alpha_{s}(M_{Z})$
 prediction.
The error bar on the $\alpha_{s}(M_{Z})$ range extracted from
 experiment (dash-dot line) and the two-loop contribution to the
 $\alpha_{s}(M_{Z})$ prediction (dotted line) are given for comparison.}
 \end{figure}

 \begin{figure}
\caption{The $m_{t}^{pole} - \tan\beta$
plane is divided into five different regions.
Two areas (low- and high-$\tan\beta$ branches)
are consistent with perturbative
Yukawa unification ($h_{b} = h_{\tau}$ at $M_{G}$)
and with $0.85m_{b}^{0}(5\,\GeV) < 4.45$ GeV.
Between the two branches the $b$-quark
mass is too high. For a too low (high) $\tan\beta$, $h_{t}$
($h_{b}$)
diverges. The strip where all three (third-family)
Yukawa couplings unify intersects
the allowed high-$\tan\beta$ branch
and is indicated as well (dash-dot line).
Corrections to the $h_{t}/h_{b}$ ratio induce a $\sim \pm 5\%$ uncertainty
in the $m_{t}^{pole}$ range that corresponds
to each of the points in the three-Yukawa strip.
$\alpha_{s}(M_{Z})$, $\alpha_{G}$, and the unification scale
used in the calculation are the ones
predicted by the MSSM coupling constant unification, and
are sensitive to the $t$-quark pole mass, $m_{t}^{pole}$.
The $m_{t}^{pole}$  range
suggested by the electroweak data is indicated
(dashed lines) for comparison.
$m_{t}^{pole}$   is in GeV.}
\end{figure}

\end{document}